\documentclass[12pt]{article}
\usepackage{epsfig}
\usepackage{epstopdf}
\usepackage{amsfonts}
\usepackage{amsmath}
\usepackage{amssymb}
\usepackage{graphicx}
\usepackage[colorlinks=true,linkcolor=red,citecolor=blue]{hyperref}
\begin{document}
\title{Comprehensive Study of $\overline B^0\to K^0(\overline K^0) K^\mp\pi^\pm$ Decays in the Factorization Approach}
\author{Ying Li\footnote{Email:liying@ytu.edu.cn}\\
\small {\it Department of Physics, Yantai University, Yantai 264-005, China } \\}
\maketitle
\begin{abstract}
Using the factorization approach, we investigate the $\overline B^0
\to \overline K^0K^+\pi^-$ and $\overline B^0 \to K^0K^-\pi^+$
decays individually including the resonant and nonresonant
contributions. Under the flavor SU(3) symmetry, we obtain the total
branching fraction $BR [\overline B^0 \to (\overline
K^0K^+\pi^-+K^0K^-\pi^+)]=
(7.17^{+0.50+1.97+0.08}_{-0.53-1.42-0.07})\times 10^{-6}$, which is
in agreement with the recent measurements of BaBar and LHCb within
errors. For the decay  $\overline B^0 \to \overline K^0K^+\pi^-$,
the nonresonant background and $a_0^+(1450)$ pole in the
current-induced process provide large contribution, the latter of
which has not been included in previous studies. On the contrary,
the decay $\overline B^0 \to K^0K^-\pi^+$ is dominated by the
nonresonant background and the offshell $\rho^-$ pole. When the
flavor SU(3) symmetry breaking and the final-state rescattering are
considered under two different scenarios, the results can also
accommodate the experimental data with large uncertainties.
Moreover, the direct $CP$ asymmetry of $\overline B^0 \to \overline
K^0K^+\pi^-$ is found to be sensitive to the matrix element of
scalar density. These predictions could be further tested in the
LHCb experiment or Super-b factory in future.
\end{abstract}
\newpage
Over past few years, more attention has been paid to charmless
three-body $B$ decays both experimentally and theoretically, because
by studying them one is allowed to extract the
Cabibbo-Kobayashi-Maskawa (CKM) angles, probe the sources of $CP$
violation and even search for the possible effects of new physics
beyond the standard model. In addition, the three-body decays of $B$
meson can help us study two-body decays involving vector or scalar
mesons, because most vector and scalar mesons are unstable and decay
to two pseudoscalar particles. On the experimental side, some
three-body decays of $B$ meson have been measured by Belle, BaBar
and LHCb experiment \cite{Amhis:2012bh}. The theoretical activity
has run in parallel, and certain works, within different approaches
such as factorization approach \cite{Deshpande:1995nu, Cheng:2002qu,
Cheng:2013dua}, diagrammatic approach combined with $\mathrm{SU}(3)$
symmetry \cite{Lorier:2010xf,Bhattacharya:2013cvn} and perturbative
QCD approach \cite{Chen:2002th}, have been advocated to study the $B
\to PPP$ decays.

Recently, LHCb collaboration updated the previous measurements of
branching fractions of $B \to KK\pi$ \cite {LHCb:2012pja} and the
latest result $BR [\overline B^0 \to (\overline K^0K^+\pi^- +
K^0K^-\pi^+)] =(6.4\pm0.9 \pm0.4 \pm0.3) \times 10^{-6}$
\cite{Aaij:2013uta} is consistent with the BaBar's result $(6.4 \pm
1.0 \pm 0.6) \times 10^{-6}$ released four years ago
\cite{delAmoSanchez:2010ur}. Theoretically, these two decays have
been analyzed by Cheng {\it et al.} recently in \cite{Cheng:2013dua}
based on the factorization approach. Without eliminating the
interference effect between these two decays, the predicted
branching fraction $(6.2^{+2.7}_{-1.7}) \times 10^{-6}$ was in good
agreement with above data. Very recently, these results have been
updated in \cite{Cheng:2014uga}, and the branching fraction is
$(4.7^{+2.8}_{-1.8}) \times 10^{-6}$ after correcting the typos in
the computer code.

In this work, we would like to reexamine in detail the decays
$\overline B^0 \to \overline K^0K^+\pi^-$ and $\overline B^0 \to
K^0K^-\pi^+$ in the factorization approach for the following
reasons: (i) As pointed out in \cite{Cheng:2014uga}, there are some
typos in the computer code of \cite{Cheng:2013dua}, and the
contributions of $\rho$ and $a_0(1450)$ poles have not been
included. Although the contributions for the aforementioned poles
have been added in the decay $\overline B^0\to K^0K^-\pi^+$ in
\cite{Cheng:2014uga}, their contributions have been omitted in the
decay $\overline B^0 \to \overline K^0K^+\pi^-$. Our results in the
following show that $\rho^+$ pole dominates the resonant
contribution. (ii) The results of three-body decays $B^-\to
K^+K^-K^-$ \cite{Cheng:2013dua} and $\overline B_s^0\to K_S
K^\mp\pi^\pm$   \cite{li:2014} show us that the flavor SU(3) symmetry
violation and the final-state rescattering may be large in three-body
charmless  $B$ decays. In order to describe these effects, an extra
strong phase $\delta$ \cite{Cheng:2013dua} or a phenomenological
parameter $\beta$ \cite{li:2014} has been introduced. In the present
work, we wish to examine the effects of SU(3) asymmetry and the final-state rescattering in the decays $\overline B^0 \to \overline
K^0K^+\pi^-$ and $\overline B^0 \to K^0K^-\pi^+$ individually. (iii)
The branching fractions and $CP$ asymmetries of nonresonant
backgrounds have not been given in previous studies, and we will
discuss them in detail here. (iv) We will discuss the resonant
contributions explicitly, which were incorrect in
\cite{Cheng:2013dua} and were absent in \cite{Cheng:2014uga}. (v)
The formulas, such as the individual amplitudes and the parameterized
form factors will be given explicitly.

In the factorization approach, the amplitude is usually split into
three distinct factorizable terms, the current-induced process with
a meson emission, the transition process and the annihilation
process, though the factorization in $B$ meson three-body decay has
not been proven. In the practical calculations, the contributions
from the annihilations are assumed to be power suppressed and
ignored. For the calculations of the nonresonant contributions to
$\langle KK(\pi)|(\bar q b)_{V-A}|B\rangle$ in the current-induced
processes, we extend the results obtained in \cite{Cheng:2013dua}
where most experimental results were successfully reproduced with
heavy meson chiral perturbative theory (HMChPT)\cite{Yan:1992gz}
that combines the heavy quark symmetry and the chiral Lagrangian
approach, although applicability of this framework in all kinematics
region is still controversial \cite{Meissner:2013hya}. In the $B$
meson decays, heavy quark symmetry is expected to be even better,
while the chiral perturbative theory might be less reliable due to
large energies of the light mesons in the final state. It is accepted
that the HMChPT is valid at small recoil momentum, which means that
the HMChPT could be applied in a small fraction of the whole Dalitz
plot; nevertheless, it is not justified to apply it to a certain
kinematic region and then generalize it to the region beyond its
validity. To overcome this shortcoming, Cheng {\it et al} proposed
in \cite{Cheng:2013dua} that the momentum dependence of nonresonant
amplitudes is in an exponential form $\exp [-\alpha_{\rm NR}p_B\cdot
(p_i+p_j)]$ so that the HMChPT results are recovered in the soft
meson limit, and the mode independence parameter $\alpha_{\rm
NR}=0.081^{+0.015}_{-0.009}\mathrm{GeV}^{-2}$ has been fixed from
the decay of $B^-\to \pi^+\pi^-\pi^-$, since it is dominated by the
nonresonant contribution.

The effective Hamiltonian for the process $b \to d\bar q q$ is given by
\begin{eqnarray} \label{eq:Tp}
&&H_{\rm eff}=\frac{G_F}{\sqrt2}\sum_{p=u,c}V_{pb}V_{pd}^*\left\{
 a_1 \delta_{pu} (\bar u b)_{V-A}\otimes(\bar d u)_{V-A}
 +a_2 \delta_{pu} (\bar d b)_{V-A}\otimes(\bar u u)_{V-A}\right.
 \nonumber\\
 &&
 +a_3(\bar d b)_{V-A}\otimes\sum_q(\bar q q)_{V-A}
+a^p_4\sum_q(\bar q b)_{V-A}\otimes(\bar d q)_{V-A} +a_5(\bar d b)_{V-A}\otimes\sum_q(\bar q q)_{V+A}\nonumber\\
 &&
 -2 a^p_6\sum_q(\bar q b)_{S-P}\otimes(\bar d q)_{S+P}
 +a_7(\bar d b)_{V-A}\otimes\sum_q\frac{3}{2} e_q (\bar q q)_{V+A}
 -2a^p_8\sum_q(\bar q b)_{S-P}\otimes\frac{3}{2} e_q (\bar d q)_{S+P}\nonumber\\
 &&\left.
 +a_9(\bar d b)_{V-A}\otimes\sum_q\frac{3}{2}e_q (\bar q q)_{V-A}+a^p_{10}\sum_q(\bar q b)_{V-A}\otimes\frac{3}{2}e_q(\bar d
 q)_{V-A}\right \},
\end{eqnarray}
where $(\bar q_1q_2)_{V\pm A}=\bar q_1 \gamma_\mu(1\pm\gamma_5)q_2$
and $(\bar q_1q_2)_{S\pm P}=\bar q_1 (1\pm\gamma_5)q_2$. The
corresponding effective Wilson coefficients at the renormalization
scale $\mu=2.1$~GeV are listed as \cite{Cheng:2013dua}
\begin{eqnarray} \label{eq:ai}
 &&
 a_1\approx0.99+0.037 i,\quad
 a_2\approx 0.19-0.11i, \quad
 a_3\approx -0.002+0.004i, \quad
 a_5\approx0.0054-0.005i,  \nonumber \\
 &&
 a_4^u\approx -0.03-0.02i, \quad
 a_4^c\approx-0.04-0.008i,\quad
 a_6^u\approx -0.06-0.02i, \quad
 a_6^c\approx -0.06-0.006i,\nonumber\\
 &&
 a_7\approx 0.54\times 10^{-4} i,\quad
 a_8^u\approx (4.5-0.5i)\times10^{-4},\quad
 a_8^c\approx (4.4-0.3i)\times10^{-4},   \nonumber\\
 &&
 a_9\approx -0.010-0.0002i,\quad
 a_{10}^u \approx (-58.3+ 86.1 i)\times10^{-5},\quad
 a_{10}^c \approx (-60.3 + 88.8 i)\times10^{-5},
\end{eqnarray}
where the strong phases arise from vertex corrections and penguin contractions.

According to the effective Hamiltonian, the factorizable amplitudes
for the $\overline B^0\to K^0 K^-\pi^+$ and $\overline B^0\to
\overline K^0 K^+\pi^-$ can be respectively read
\begin{eqnarray} \label{eq:amp1}
&&A(\overline B^0 \to \overline  K^0 K^+\pi^-)\nonumber \\&=&\frac{G_F}{\sqrt2}\sum_{p=u,c}V_{pb}V_{pd}^*
\left\{\langle K^+ \overline K^0|(\bar u b)_{V-A}|\overline B^0\rangle \langle \pi^-|(\bar d u)_{V-A}|0\rangle
\left[a_1 \delta_{pu}+a^p_4+a_{10}^p-(a^p_6+a^p_8) r_\chi^\pi\right] \right.\nonumber\\
&&+\langle \overline K^0|(\bar s b)_{V-A}|\overline B^0\rangle
\langle K^+\pi^-|(\bar d s)_{V-A}|0\rangle
\Big [a^p_4-\frac{1}{2}a^p_{10}\Big]  \nonumber\\
&& + \langle \overline K^0|\bar s b|\overline B^0\rangle \langle K^+\pi^-|\bar d s|0\rangle
     \Big[-2 a^p_6+a^p_8 \Big]  \nonumber\\
&&  + \langle  \overline K^0 K^+\pi^-|(\bar uu)_{V-A}|0\rangle\langle 0|(\bar d b)_{V-A}|\overline B^0\rangle
       \Big[a_2\delta_{pu}+a_3+a_9\Big]\nonumber\\
&&  + \langle  \overline K^0 K^+\pi^-|(\bar dd)_{V-A}|0\rangle\langle 0|(\bar d b)_{V-A}|\overline B^0\rangle
       \Big[a_3+a_4^p-\frac{1}{2}a_9-\frac{1}{2}a_{10}^p\Big]\nonumber\\
&&  + \langle  \overline K^0 K^+\pi^-|(\bar ss)_{V-A}|0\rangle\langle 0|(\bar d b)_{V-A}|\overline B^0\rangle
       \Big[a_3-\frac{1}{2}a_9\Big]\nonumber\\
&&  \left.+  \langle \overline K^0 K^+\pi^-|\bar d(1+\gamma_5) d|0\rangle
      \langle 0|\bar d\gamma_5 b|\overline B^0\rangle
      \Big[ 2a^p_6-a^p_8\Big]
\right\},
\end{eqnarray}
\begin{eqnarray} \label{eq:amp2}
&&A(\overline B^0 \to K^0 K^-\pi^+)\nonumber\\&=&\frac{G_F}{\sqrt2}\sum_{p=u,c}V_{pb}V_{pd}^*
\left\{\langle K^-\pi^+|(\bar s b)_{V-A}|\overline B^0\rangle \langle K^0|(\bar d s)_{V-A}|0\rangle
     \Big[a^p_4-\frac{1}{2}a^p_{10}-(a^p_6-\frac{1}{2}a^p_8) r_\chi^K\Big] \right.\nonumber\\
&& +\langle \pi^+|(\bar u b)_{V-A}|\overline B ^0\rangle \langle K^- K^0|(\bar d u)_{V-A}|0\rangle
     \Big [a_1\delta_{pu}+a_4^p+a_{10}^p\Big ]  \nonumber\\
&& + \langle\pi^+|\bar u b|\overline B ^0\rangle\langle K^- K^0|\bar d u|0\rangle
     \Big[-2a^p_6-2a^p_8\Big] \nonumber\\
&&  + \langle  K^0 K^-\pi^+|(\bar uu)_{V-A}|0\rangle\langle 0|(\bar d b)_{V-A}|\overline B^0\rangle
       \Big[a_2\delta_{pu}+a_3+a_9\Big]\nonumber\\
&&  + \langle  K^0 K^-\pi^+|(\bar dd)_{V-A}|0\rangle\langle 0|(\bar d b)_{V-A}|\overline B^0\rangle
       \Big[a_3+a_4^p-\frac{1}{2}a_9-\frac{1}{2}a_{10}^p\Big]\nonumber\\
&&  + \langle  K^0 K^-\pi^+|(\bar ss)_{V-A}|0\rangle\langle 0|(\bar d b)_{V-A}|\overline B^0\rangle
       \Big[a_3-\frac{1}{2}a_9\Big]\nonumber\\
&&  + \left. \langle K^0 K^-\pi^+|\bar d(1+\gamma_5) d|0\rangle
      \langle 0|\bar d\gamma_5 b|\overline B^0\rangle
      \Big[ 2a^p_6-a^p_8\Big]
\right\}.
\end{eqnarray}
In each amplitude, the last four terms arising from the annihilation
contributions will be ignored  in the following since they are power
suppressed and also $\alpha_s$ suppressed.

As mentioned before, in order to study the nonresonant background of
the matrix element $\langle M_1(p_1)M_2(p_2)|(\bar d
b)_{V-A}|\overline B^0\rangle$, where the meson $M_1$ involves the
spectator quark $\bar d$, we shall use the HMChPT and generalize the
results obtained previously in calculating the decay $B^- \to
\pi^+\pi^-\pi^-$ in \cite{Cheng:2013dua}. We then obtain the matrix
element, for example $\langle \overline K^0(p_1)K^+(p_2)|(\bar u
b)_{V-A}|\overline B^0\rangle$,  in the general form
\begin{eqnarray} \label{eq:romegah}
  &&\langle  \overline K^0(p_1)K^+(p_2)|(\bar u b)_{V-A}|\overline B^0\rangle \nonumber \\
 &&=i r(p_B-p_1-p_2)_\mu+i\omega_+(p_2+p_1)_\mu+i\omega_-(p_2-p_1)_\mu
 +h\,\epsilon_{\mu\nu\alpha\beta}p_B^\nu (p_2+p_1)^\alpha
 (p_2-p_1)^\beta.
\end{eqnarray}
The form factors $\omega_{\pm}$ and $r$ can be evaluated within the HMChPT and given by
\begin{eqnarray} \label{eq:r&omega}
\omega_+ &=& -{g\over f_\pi^2}\,{f_{B^*}m_{B^*}\sqrt{m_Bm_{B^*}}\over
 s_{23}-m_{B^*}^2}\left[1-{(p_B-p_1)\cdot p_1\over
 m_{B^*}^2}\right]+{f_B\over 2f_\pi^2},   \\
 \omega_- &=& {g\over f_\pi^2}\,{f_{B^*}m_{B^*}\sqrt{m_Bm_{B^*}}\over
 s_{23}-m_{B^*}^2}\left[1+{(p_B-p_1)\cdot p_1\over
 m_{B^*}^2}\right],   \\
 r &=& {f_B\over 2f_\pi^2}-{f_B\over
 f_\pi^2}\,{p_B\cdot(p_2-p_1)\over
 (p_B-p_1-p_2)^2-m_B^2}+{2gf_{B^*}\over f_\pi^2}\sqrt{m_B\over
 m_{B^*}}\,{(p_B-p_1)\cdot p_1\over s_{23}-m_{B^*}^2} \nonumber \\
 && -{4g^2f_B\over f_\pi^2}\,{m_Bm_{B^*}\over
 (p_B-p_1-p_2)^2-m_B^2}\,{p_1\!\cdot\!p_2-p_1\!\cdot\!(p_B-p_1)\,p_2\!\cdot\!
 (p_B-p_1)/m_{B^*}^2 \over s_{23}-m_{B^*}^2 },
\end{eqnarray}
where $s_{ij}\equiv (p_i+p_j)^2$. The heavy-flavor independent
strong coupling $g$ has been extracted from the $D^{*+}$ decay
width, $g=-0.59\pm0.01\pm0.07$. Together with the aforementioned
exponential form, we obtain the nonresonant amplitude of
current-induced process as
\begin{eqnarray} \label{eq:AHMChPT}
&& \langle K^+ \overline K^0|(\bar u b)_{V-A}| \overline B^0\rangle^{\rm NR} \langle \pi^-|(\bar du)_{V-A}|0\rangle\nonumber\\
&&= -\frac{f_\pi}{2}\left[2 m_3^2 r+(m_B^2-s_{12}-m_3^2) \omega_+
+(s_{23}-s_{13}-m_2^2+m_1^2) \omega_-\right]\,e^{-\alpha_{_{\rm NR}}
p_B\cdot(p_1+p_2)}e^{i\phi_{12}},
\end{eqnarray}
and the unknown strong phase $\phi_{12}$ will be set to zero for
simplicity. For the decay $\overline B^0 \to K^0 K^-\pi^+$, the expression of $\langle K^- \pi^+|(\bar s b)_{V-A}|\overline B^0\rangle^{\rm NR}\langle K^0|(\bar ds)_{V-A}|0\rangle$ is similar to  Eqs.
(\ref{eq:r&omega}-\ref{eq:AHMChPT}).

For the resonant contribution of the current-induced process, the
scalar $a_0^+(1450)$ and vector $\rho^+$ contribute to the  matrix
element $\langle K^+ \overline K^0|(\bar u b)_{V-A}|\overline
B^0\rangle$, which can be written as:
\begin{eqnarray}
&& \langle K^+ \overline K^0|(\bar u b)_{V-A}|\overline B^0\rangle^R\langle \pi^-|(\bar d u)_{V-A}|0\rangle \nonumber \\
&=&{-f_\pi}\,{g^{\rho^{+}\to K^+\overline K^0}\over
s_{23}-m_{\rho}^2+im_{\rho}\Gamma_{\rho}}\left\{\left(s_{13}-s_{12}\right)\Big[
m_{\rho}A_0^{B\rho}(q^2)
+\frac{A_2^{B\rho}(q^2)}{2(m_B+m_{\rho})}(s_{23}-m_{\rho}^2) \Big]
\right\}\nonumber\\&& + f_\pi{g^{a_0^+(1450)\to K^+ \overline
K^0}\over s_{23}-
m_{a_0(1450)}^2+im_{a_0(1450)}\Gamma_{a_0(1450)}}\Big[(m_B^2-m_{a_0(1450)}^2)F_0^{Ba_0(1450)}(q^2)
\nonumber\\&& +(m_{a_0(1450)}^2-s_{23})F_1^{Ba_0(1450)}(q^2)\Big].
\end{eqnarray}
Here the resonant effects are described in terms of the usual
Breit-Wigner formalism. Similarly, the matrix element  $\langle K^-
\pi^+|(\bar s b)_{V-A}|\overline B^0\rangle$ receives contributions
of vector meson $\overline K^{*0}$ and scalar $\overline
K_0^{*0}(1430)$, and the expression is given as
\begin{eqnarray}
 && \langle K^-(p_2)\pi^+(p_1)|(\bar sb)_{V-A}|\overline B^0\rangle^R ~\langle K^0(p_3)|(\bar
ds)_{V-A}|0\rangle \nonumber \\
&=&{-f_K}\,{g^{K^*\to K^-\pi^+}\over
s_{12}-m_{K^*}^2+im_{K^*}\Gamma_{K^*}}\left\{\Big[s_{13}-s_{23}+\frac{(m_B^2-m_K^2)(m_K^2-m_\pi^2)}
{m_{K^*}^2}\Big]\Big[ m_{K^*}A_0^{BK^*}(q^2)\right.\nonumber
\\&&+\frac{A_2^{BK^*}(q^2)}{2(m_B+m_{K^*})} (s_{12}-m_{K^*}^2)
\Big]+\left (m_K^2-m_\pi^2\right)\left
(1-\frac{s_{12}}{m_{K^*}^2}\right)\nonumber \Big[
m_{K^*}A_0^{BK^*}(q^2) \\&&\left.-(m_B+m_{K^*})A_0^{BK^*}(q^2)
+\frac{A_2^{BK^*}(q^2)}{2(m_B+m_{K^*})}(s_{12}-m_{K^*}^2)
\Big]\right\}\nonumber\\&& + f_K{g^{K_{0}^*\to K^-\pi^+}\over
s_{12}-
m_{K_{0}^*}^2+im_{K_{0}^*}\Gamma_{K_{0}^*}}\Big[(m_B^2-m_{K_{0}^*}^2)F_0^{BK_0^*}(q^2)
+(m_{K_{0}^*}^2-s_{12})F_1^{BK_0^*}(q^2)\Big].
\end{eqnarray}

Now, we shall evaluate the transition processes. The matrix element
$\langle K^+(p_2)\pi^-(p_3)|(\bar d s)_{V-A}|0\rangle$ can be
parameterized as:
\begin{eqnarray}\label{eq:FKPI}
 &&\langle \overline K^0(p_1)|(\bar s b)_{V-A}|\overline B^0\rangle \langle K^+(p_2)\pi^-(p_3)|(\bar d s)_{V-A}|0\rangle\nonumber\\
&=& -F_1^{BK}(s_{23})F_1^{K\pi}(s_{23})\left[s_{13}-s_{12}-{(m_B^2-m_K^2)(m_K^2-m_\pi^2)
\over s_{23}}\right]\nonumber\\&&  -F_0^{BK}(s_{23})F_0^{K\pi}(s_{23}){(m_B^2-m_K^2)(m_K^2-m_\pi^2)\over s_{23}}.
\end{eqnarray}
A recent detailed analysis of $B^- \to K^-\pi^+\pi^- $ decay in
\cite{Cheng:2013dua} indicates that the nonresonant contribution
(weak form factor $F^{K\pi}(q^2)$) plays negligible role, so it can
be ignored safely.  The contributions from vector and scalar poles
to the form factors have the expressions as:
 \begin{eqnarray} \label{eq:F1Kpi}
 F^{K\pi,R}_{1}(s) &=& {m_{K^*}f_{K^*}g^{K^*\to K\pi}\over
 m_{K^*}^2-s-im_{K^*}\Gamma_{K^*}},   \\\label{eq:F1Kpi2}
F^{K\pi,R}_0(s) &=&{f_{K_{0}^*}g^{K_{0}^*\to K\pi}\over
 m_{K_{0}^*}^2-s-im_{K_{0}^*}\Gamma_{K_{0}^*}}\,{s\over m_K^2-m_\pi^2}
 -{m_{K^*}f_{K^*}g^{K^*\to K\pi}\over
 m_{K^*}^2-s-im_{K^*}\Gamma_{K^*}}\big[-1+ {s\over m^2_{K^*}}\Big].
\end{eqnarray}
Note that for the scalar meson, the scale-dependent scalar decay
constant $\bar f_S$ and the vector decay constant $f_S$ are defined
by
\begin{eqnarray} \label{eq:decayc}
\langle S|\bar q_2q_1|0\rangle=m_S\bar f_S,\,\,\,\langle S(p)|\bar q_2\gamma_\mu q_1|0\rangle=f_S p_\mu.
\end{eqnarray}
The two decay constants are related by the equation of motion
\begin{eqnarray}  \label{eq:EOM}
 \mu_Sf_S=\bar f_S, \qquad\quad{\rm with}~~\mu_S={m_S\over m_2(\mu)-m_1(\mu)},
\end{eqnarray}
where $m_2$ and $m_1$ are the running current quark masses and $m_S$
is the scalar meson mass.

For the term $\langle\pi^+|(\bar u b)_{V-A}|\overline B
^0\rangle\langle K^- K^0|(\bar d u)_{V-A}|0\rangle $, the
contributions are not only from nonresonant contribution $\langle
K^- K^0|(\bar d u)_{V-A}|0\rangle $ that is related to $\langle K^-
K^+|(\bar u u)_{V-A}|0\rangle $ via flavor SU(3) symmetry, but also
from the offshell vector meson $\rho^-$. For the expression
for the nonresonant background and its inner functions we refer to \cite{Cheng:2013dua};
the contributions from resonant particles are similar to Eqs.(\ref{eq:FKPI}-\ref{eq:F1Kpi2}).

We also need to specify the amplitudes induced from the scalar
densities. With the equation of the motion, we are led to, for
example,
\begin{eqnarray}
\langle \overline K^0(p_1)|\bar s b|\overline B^0(p_B)= \frac{m_B^2-m_K^2}{m_b-m_s} F_0^{BK}(s_{23}).
\end{eqnarray}
The matrix element $\langle K^+\pi^-|\bar d s|0\rangle$ receives both
resonant and nonresonant contributions:
\begin{eqnarray}
 \langle K^+(p_2) \pi^-(p_3)|\bar ds|0\rangle
 =\frac{m_{{K^*_0}} \bar f_{{K^*_0}} g^{{K^*_0}\to K^+\pi^-}}{m_{{K^*_0}}^2-s_{23}-i
 m_{{K^*_0}}\Gamma_{{K^*_0}}}+\langle K^+(p_2) \pi^-(p_3)|\bar s
 d|0\rangle^{\rm NR}.
\end{eqnarray}
The unknown  scalar density $\langle K^+\pi^-|\bar ds|0\rangle^{\rm NR}$
is related to $\langle K^+K^-|\bar ss|0\rangle$ via the SU(3)
symmetry, e.g.
\begin{eqnarray} \label{eq:Kpim.e.SU3}
 \langle K^+(p_2) \pi^-(p_3)|\bar ds|0\rangle^{\rm NR}=\langle K^+(p_2)K^-(p_3)|\bar
 ss|0\rangle^{\rm NR}=f_s^{\rm NR}(s_{23}),
\end{eqnarray}
with the expression as
\begin{eqnarray}\label{fNR}
f_s^{\rm NR}&=&\frac{v}{3}(3 F_{\rm NR}+2F'_{\rm NR})+\sigma_{_{\rm NR}}
 e^{-\alpha s_{23}}\,\,\,  {\rm and} \,\,\, v=\frac{m_{K^+}^2}{m_u+m_s}=\frac{m_K^2-m_\pi^2}{m_s-m_d}.
\end{eqnarray}
We will use the experimental measurement $\alpha=(0.14\pm0.02)\mathrm{GeV}^{-2}$ \cite{BaBarKpKmK0}.
The functions $F^{(\prime)}_{\rm NR}(s)$ is given in \cite{Cheng:2002qu} by
\begin{eqnarray}
 F^{(\prime)}_{\rm NR}(s_{23})=\left(\frac{x^{(\prime)}_1}{s_{23}}
 +\frac{x^{(\prime)}_2}{s_{23}^2}\right)
 \left[\ln\left(\frac{s_{23}}{\tilde\Lambda^2}\right)\right]^{-1},
\end{eqnarray}
with $\tilde\Lambda\approx 0.3$ GeV, $x_1=-3.26~{\rm GeV}^2,
x_2=5.02~{\rm GeV}^4, x'_1=0.47~{\rm GeV}^2$ and $x'_2=0$. The
expression for the nonresonant form factor is motivated by the
asymptotic constraint from perturbative QCD, namely,
$F(t)\to(1/t)\times [\ln(t/\tilde{\Lambda^2})]^{-1}$ in the large
$t$ limit \cite{Brodsky}. For the unknown parameter $\sigma_{_{\rm
NR}}$, we firstly adopt the value fitted from $B \to KKK$ decays,
$\sigma_{_{\rm NR}}= e^{i\pi/4} \left(3.39^{+0.18}_{-0.21}\right)
\,{\rm GeV}$, which is called scenario-1 (S1). However, in
\cite{Cheng:2013dua}, it is found that the predicted $A_{CP}(B^-\to
K^-\pi^+\pi^-)$ and $A_{CP}(B^-\to K^+K^-\pi^-)$ with the flavor
SU(3) symmetry are wrong in signs confronting the corresponding
experimental data, which implies that the flavor SU(3) symmetry
violation and the final-state rescattering might be important.  To
pick up these kinds of the contributions, two phenomenological
ways have been adopted. In \cite{Cheng:2013dua}, an extra
strong phase $\delta$ has been introduced in Eq.(\ref{fNR}) and its
value was fixed to be $\pi$ by the $CP$ asymmetries of $B \to KK\pi$
and $B\to K\pi\pi$, and the expression is then given by
\begin{eqnarray}
 \langle K^+(p_2)\pi^-(p_3)|\bar d s|0\rangle^{\rm NR}
 &\approx & \frac{v}{3}(3 F_{\rm NR}+2F'_{\rm NR})+\sigma_{_{\rm NR}}
 e^{-\alpha s_{23}}e^{i\pi}.
\end{eqnarray}
In \cite{li:2014}, a phenomenological coefficient $\beta=0.7 \pm
0.2$ has been proposed to describe these effects,
\begin{eqnarray}
 \langle K^+(p_2)\pi^-(p_3)|\bar d s|0\rangle^{\rm NR}
 &\approx & \beta \langle K^+(p_2)K^-(p_3)|\bar s s|0\rangle^{\rm NR}
\end{eqnarray}
by which the experimental data of $B_s\to K_s K^\pm \pi^\mp$ can be
reproduced \footnote {The experimental data of $A_{CP}(B^-\to
K^-\pi^+\pi^-)$, $A_{CP}(B^-\to K^+K^-\pi^-)$ and $B_s\to K_s K^\pm
\pi^\mp$ cannot be explained under above scenarios simultaneously.}.
For convenience, we name above two ways as scenario-2 (S2) and
scenario-3 (S3). Under S2, the absolute value  $|\sigma_{_{\rm
NR}} e^{-\alpha s_{23}}|$ is unchanged with flipping the relative
sign between real part and imaginary part; while under S3, the strong
phase is kept with decreasing the absolute value.

In fact, the scalar $K\pi$ form factor has been attached more attentions
because it plays important role in studying $B \to K^{(*)}
\ell^+\ell^-$ decays \cite{Meissner:2013pba}. In the past years, it has been calculated
within a variety of approaches such as the chiral perturbation theory \cite{Meissner:2000bc} and
the dispersion relations \cite{Jamin:2000wn}, but it is still difficult to extract it in the
experimental side due to the mixing of $S$ and $P$ waves. If the
aforementioned scenarios could be discriminated in the future, it might give us some hints in studying this scalar form factor.

For the decay mode $\overline B^0 \to K^0 K^-\pi^+$, the
contribution of  $\langle K^0K^-|\bar d u|0\rangle$ arising from
nonresonant background  is similar to  $\langle K^+K^-|\bar s
s|0\rangle$ according to the SU(3) symmetry.  In principle, the
flavor SU(3) asymmetry should be included, however previous
studies show that its effect is small enough to be ignored. Together
with the resonant contribution of $a_0^-(1450)$, we are led to
\begin{eqnarray}
 \langle K^-(p_2) K^0(p_3)|\bar du|0\rangle
 =\frac{m_{a_0^-(1450)} \bar f_ {a_0^-(1450)} g^{a_0^-(1450)\to K^-K^0}}{m_{{a_0^-(1450)}}^2-s_{23}-i
 m_{{a_0^-(1450)}}\Gamma_{{a_0^-(1450)}}}+\frac{v}{3}(3 F_{\rm NR}+2F'_{\rm NR})+\sigma_{_{\rm NR}}
 e^{-\alpha s_{23}}.
\end{eqnarray}

\begin{table}[t]
\caption{Predicted branching fractions (in units of $10^{-6}$) of
resonant and nonresonant (NR) contributions to $\overline
B^0\to\overline K^0 K^+\pi^-$ and $K^0K^-\pi^+$ under S1. }
\begin{center} \label{tab:results}
\begin{tabular}{c  c |c c }\hline\hline
\multicolumn{2}{c |}{$\overline B^0\to
\overline K^0 K^+\pi^-$} &\multicolumn{2}{c}{$\overline B^0\to  K^0 K^-\pi^+$}  \\ \hline
$K^{*0}\overline K^0$
 & $0.24^{+0.00+0.04+0.01}_{-0.00-0.04-0.01}$
 & $ \overline K^{*0} K^0$
 &  $0.05^{+0.00+0.11+0.00}_{-0.00-0.04-0.00}$\\
$K_0^{*0}(1430)\overline K^0$
 & $0.94^{+0.00+0.19+0.03}_{-0.00-0.17-0.02}$
 & $ \overline K_0^{*0}(1430) K^0$
 & $0.02^{+0.00+0.06+0.00}_{-0.00-0.02-0.00}$  \\
$a_0^{+}(1450) \pi^-$
 & $2.04^{+0.00+0.54+0.02}_{-0.00-0.44-0.01}$
 & $a_0^{-}(1450) \pi^+$
 & $0.20^{+0.00+0.06+0.01}_{-0.00-0.05-0.01}$
 \\
$\rho^{+} \pi^-$
 & $0.29^{+0.00+0.06+0.00}_{-0.00-0.06-0.01}$
 & $\rho^{-} \pi^+$
 & $0.48^{+0.00+0.12+0.01}_{-0.00-0.11-0.01}$ \\\hline
NR
 &$2.71^{+0.57+0.64+0.04}_{-0.64-0.37-0.04}$
 &NR
 & $0.34^{+0.01+0.14+0.01}_{-0.01-0.11-0.01}$  \\
Total
 &$6.35^{+0.49+1.59+0.06}_{-0.52-1.18-0.06}$
 & Total
 & $0.82^{+0.01+0.38+0.02}_{-0.01-0.24-0.01}$\\
 \hline\hline
 \end{tabular}
\end{center}
\end{table}
\begin{table}[t]
\caption{The nonresonant and total branching fraction of $\overline
B^0\to \overline K^0 K^+\pi^-$ under S2 and S3. }
\begin{center} \label{tab:results2}
\begin{tabular}{c | c c   }
\hline\hline &NR&Total \\\hline S2
&$2.72^{+0.60+0.64+0.03}_{-0.67-0.36-0.03}$
&$5.83^{+0.51+1.49+0.04}_{-0.55-1.08-0.04}$\\
S3 &$2.18^{+0.58+0.85+0.03}_{-0.65-0.29-0.03}$
&$5.58^{+0.49+1.94+0.04}_{-0.53-1.24-0.05}$
\\\hline\hline
 \end{tabular}
\end{center}
\end{table}

In the numerical calculation of the branching fractions we follow
the discussions of the input parameters given in
\cite{Cheng:2013dua}. In Table.\ref{tab:results}, we present the
resonant and nonresonant contributions to the branching fractions of
the $\overline B^0\to \overline K^0 K^+\pi^-$ and $\overline B^0\to
K^0K^-\pi^+$ decays individually under S1. We also give the
nonresonant and total branching fractions of  $\overline B^0\to
\overline K^0 K^+\pi^-$ under S2 and S3 in Table.\ref{tab:results2}.
In above two tables, the first errors are from the uncertainties in
the parameter $\alpha_{_{\rm NR}}$ which governs the momentum
dependence of the nonresonant amplitude. The second ones arise from
the strange quark mass $m_s$, the form factors, the nonresonant
parameter $\sigma_{_{\rm NR}}$ and the flavor $SU(3)$ symmetry
breaking. And the last uncertainties are induced by the CKM
angle $\gamma$. Note that the ignored uncertainties arising from
the power corrections such as annihilations and hard-scattering
corrections may be sizable, but the estimation of them is beyond the
scope of the current work.

From Tables.\ref{tab:results} and \ref{tab:results2}, we get the
branching fractions of $\overline B^0\to  (\overline  K^0 K^+\pi^- +
K^0K^-\pi^+)$ under different scenarios, as given
\begin{eqnarray}
&&S1:  \left\{
\begin{array}{c}
 BR[\overline B^0\to  (\overline  K^0 K^+\pi^- + K^0K^-\pi^+)]_{\rm NR} =
 (3.05^{+0.58+0.78+0.05}_{-0.65-0.48-0.05})\times 10^{-6} \\
 BR[\overline B^0\to  (\overline  K^0 K^+\pi^- + K^0K^-\pi^+)]_{\rm Total}=
 (7.17^{+0.50+1.97+0.08}_{-0.53-1.42-0.07})\times 10^{-6}\end{array}
\right.;\\
&&S2:  \left\{
\begin{array}{c}
 BR[\overline B^0\to  (\overline  K^0 K^+\pi^- + K^0K^-\pi^+)]_{\rm NR} =
 (3.06^{+0.61+0.78+0.04}_{-0.68-0.47-0.04})\times 10^{-6} \\
 BR[\overline B^0\to  (\overline  K^0 K^+\pi^- + K^0K^-\pi^+)]_{\rm Total} =
 (6.65^{+0.52+1.87+0.06}_{-0.56-1.32-0.05})\times 10^{-6}\end{array}
\right.;\\
&&S3:  \left\{
\begin{array}{c}
 BR[\overline B^0\to  (\overline  K^0 K^+\pi^- + K^0K^-\pi^+)]_{\rm NR} =
 (2.52^{+0.59+0.99+0.04}_{-0.66-0.40-0.04})\times 10^{-6} \\
 BR[\overline B^0\to  ( \overline  K^0 K^+\pi^- + K^0K^-\pi^+)]_{\rm Total}=
 (6.40^{+0.50+2.32+0.06}_{-0.54-1.48-0.06})\times 10^{-6}  \end{array}
\right. .
\end{eqnarray}
Obviously, the total branching fractions are consistent with current
experimental data well within large uncertainties, thus it is hard
for us to discard anyone of them with current data. Because of some typos in the computer code for
Ref. \cite{Cheng:2013dua}, our results differ with them. Moreover,
the total branching fraction is much larger than those in
\cite{Cheng:2014uga} as they had missed the contributions of $\rho$
and $a_0(1450)$ poles in decay $\overline B^0\to  \overline  K^0
K^+\pi^-$, though the branching fractions of $\overline B^0\to  K^0
K^-\pi^+$  are identical. For comparison, we also calculate the
branching fractions of $\overline B^0\to  \overline  K^0 K^+\pi^-$
and $\overline B^0\to  K^0 K^-\pi^+$ without contribution of the
$\rho$ and $a_0(1450)$ poles under S2, and the results are given as:
\begin{eqnarray}
 BR(\overline B^0\to \overline  K^0 K^+\pi^-)  &=& (3.98^{+0.59+0.95+0.03}_{-0.64-0.60-0.03})\times 10^{-6} ;\\
 BR(\overline B^0\to  K^0K^-\pi^+ ) &=& (0.38^{+0.01+0.27+0.01}_{-0.01-0.15-0.01})\times 10^{-6},
\end{eqnarray}
and the branching fraction of $\overline B^0\to \overline K^0
K^+\pi^-$ is in agreement with the prediction in
\cite{Cheng:2014uga}.

\begin{figure*}[htb]
\centerline{\psfig{figure=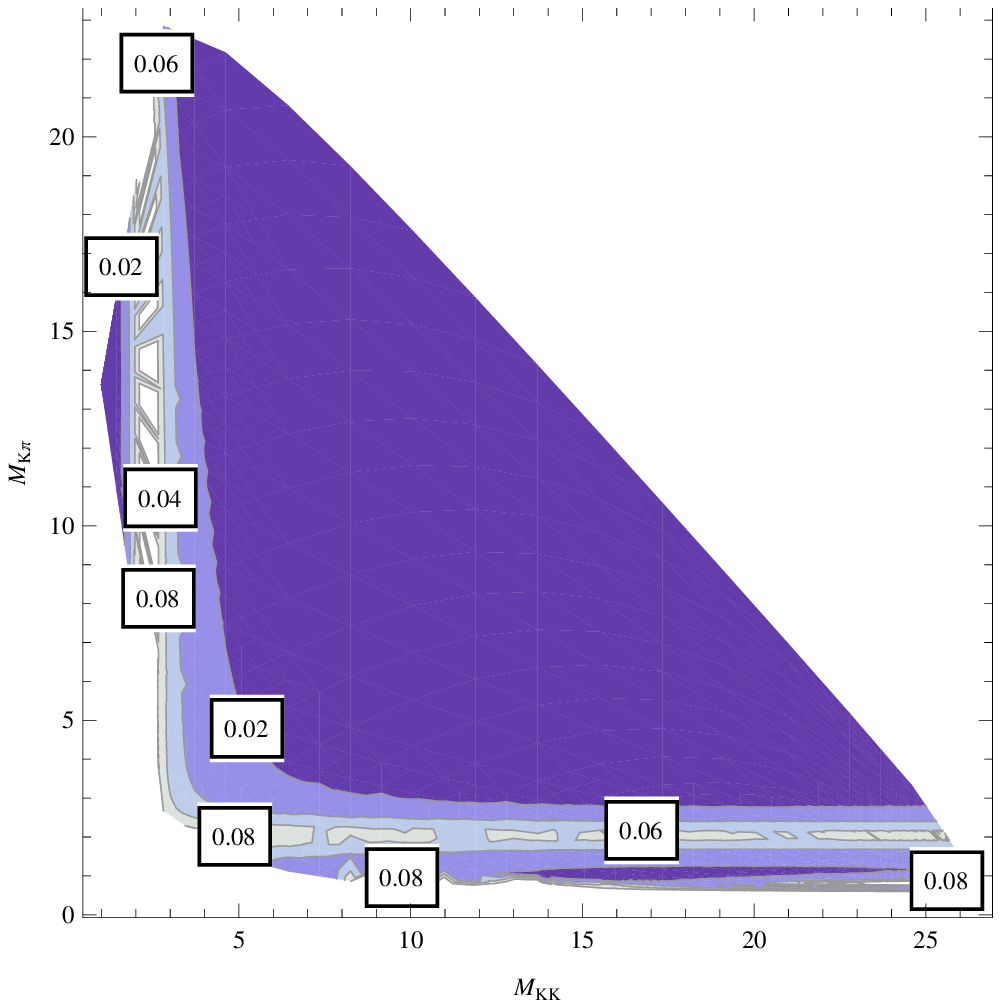,width=7cm}\,\,\,\,\,\,\,\,\,\,\,\,\,\,\,\,\,\,
\psfig{figure=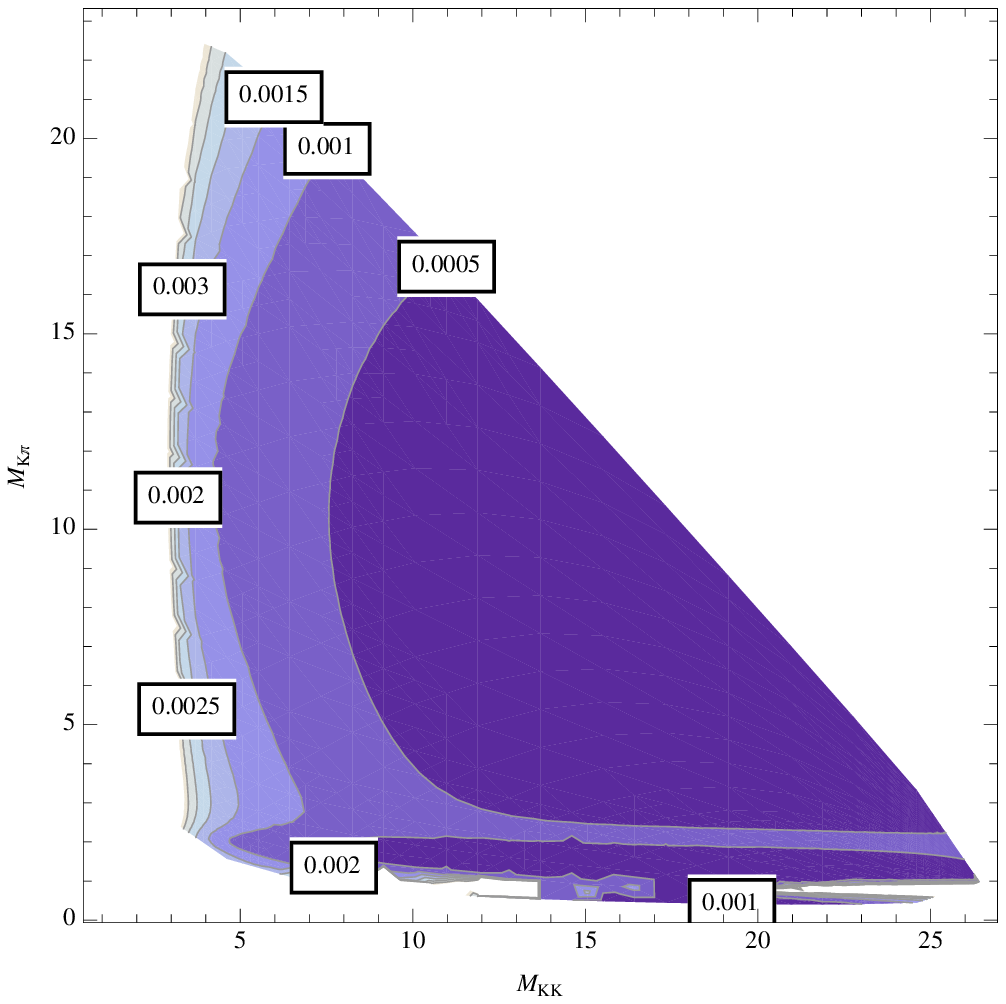,width=7cm}} \caption{Dalitz-plot
distributions of  $\overline B^0 \to \overline K^0 K^+\pi^-$ (left
panel) and $ \overline B^0 \to K^0 K^-\pi^+$ (right panel) dacays.
The $CP$ averaged differential rates are in units of $10^{-6}$
GeV$^{-4}$. }\label{fig:1}
\end{figure*}

It is obvious from Table.\ref{tab:results} that the decay $\overline
B^0\to \overline K^0 K^+\pi^-$ is dominated by the nonresonant
contribution and $a_0^+(1450)$ pole, which are of order  $42.7\%$
and $32.1\%$, respectively. As for the nonresonant background, the
current-induced process accounts for about $62\%$ due to the large
Wilson coefficient $a_1$, and the effect from the penguin suppressed
scalar density is about $38\%$. For the decay $\overline B^0 \to
K^0K^-\pi^+$, it is found that the offshell $\rho^-$ pole
plays the predominant role, though the decay $\rho^- \to K^0K^-$ is
kinematically not allowed. In these decays, the resonances $K^{*\pm}$
and $K_0^{*\pm}(1430)$ are absent because the quasi two-body decays
$\overline B^0 \to K^{*\pm} K^{\mp}$ and $\overline B^0 \to
K_0^{*\pm}(1430) K^{\mp}$ can proceed only via the pure annihilation
diagrams and they are power suppressed. In addition, the scalar
$K_0^{*0}(1430)$ and the vector $K^{*0}$ poles can only be produced
by penguins, which leads to the smaller ratios comparing to the
other contributions. The above discussions about the resonant and
nonresonant contributions are also exhibited in the Dalitz-plot
shown in Fig.~\ref{fig:1}.

Because both $\overline B^0 \to \overline K^0 K^+\pi^-$ and
$\overline B^0\to K^0K^-\pi^+$ decays have tree and penguin
contributions, their $CP$ asymmetries might be sizable. In term of
the definition of direct $CP$ violation, for example,
\begin{eqnarray}  \label{sd2}
A_{CP}=\frac{BR(\overline B^0 \to \overline K^0 K^+\pi^-)-BR(B^0 \to K^0 K^-\pi^+)}
{BR(\overline B^0 \to \overline K^0 K^+\pi^-)+BR(B^0 \to K^0 K^-\pi^+)},
\end{eqnarray}
we investigate the $CP$ asymmetries of these two decays under
different scenarios, and present the results in Table.~\ref{tab:cp}.
Note that under S2, owing to an extra strong phase $\delta=\pi$, the
sign of $CP$ asymmetry for $\overline B\to \overline K^0 K^+\pi^-$
differs from the other two cases. If both nonresonant and total $CP$
asymmetries could be studied in the LHCb experiment, it might be helpful to test
which scenario is preferable.

\begin{table}[htb]
\caption{Predicted $CP$ asymmetry ($\%$) of resonant and nonresonant
(NR) contributions to $\overline B^0\to\overline K^0 K^+\pi^-$ and
$K^0K^-\pi^+$. }
\begin{center} \label{tab:cp}
\begin{tabular}{c| c c c | c }\hline
&\multicolumn{3}{c |}{$\overline B^0\to \overline K^0
K^+\pi^-$}&$\overline B^0\to K^0K^-\pi^+$\\\hline &S1 &S2 &S3
&S1\\\hline NR &$32.5^{+0.0+2.5+0.1}_{-2.4-5.3-0.1}$
&$-25.8^{+2.0+1.7+0.1}_{-2.7-1.1-0.1}$
&$29.8^{+1.7+4.7+0.1}_{-1.9-7.2-0.1}$
&$-11.3^{+0.4+1.2+0.1}_{-0.6-1.1-0.1}$\\
Total &$14.6^{+3.1+2.7+0.1}_{-5.3-3.3-0.1}$
&$-16.9^{+1.7+1.5+0.1}_{-2.2-1.0-0.1}$
&$12.8^{+2.7+3.8+0.1}_{-4.7-3.9-0.1}$
&$-13.3^{+0.1+0.3+0.1}_{-0.1-0.6-0.1}$
\\\hline
 \end{tabular}
\end{center}
\end{table}

To summarize, we have quantitatively analyzed the resonant and
nonresonant contributions to the $\overline B^0 \to \overline K^0
K^+\pi^-$  and  $ \overline B^0 \to K^0 K^-\pi^+$  decays in the
factorization approach. Including the resonant and nonresonant
contributions, the total branching fraction $BR [\overline B^0 \to(
\overline K^0K^+\pi^- +  K^0K^-\pi^+)]= (7.17^{+0.50 +1.97+0.08
}_{-0.53 -1.42-0.07}) \times 10^{-6}$  agrees with the recent
measurements of BaBar and LHCb within errors, and the branching
fraction of $\overline B^0 \to \overline K^0K^+\pi^-$ is much larger
than that of $\overline B^0 \to K^0K^-\pi^+$. For the decay
$\overline B^0 \to \overline K^0K^+\pi^-$, the nonresonant
background and $a_0^+(1450)$ pole in the current-induced process
play important roles. On the contrary, the decay $\overline B^0 \to
K^0K^-\pi^+$ is dominated by the nonresonant background and the
offshell $\rho^-$ pole. When the flavor SU(3) asymmetry and the
final-state rescattering are involved under different scenarios, the
total branching fractions can also accommodate the experimental data
with the large uncertainties, but there are the sizable differences
among the nonresonant contributions and $CP$ asymmetries. These
predictions of $CP$ asymmetries could be further tested in the LHCb
experiment or Super-b factory in future.

\section*{Acknowledgments}
The author thanks Hai-Yang Cheng, Chun-Khiang Chua, Wei Wang and
Zhi-Tian Zou for valuable discussions and comments. This work is
supported by the National Science Foundation (Grants No. 11175151
and No. 11235005), and the Program for New Century Excellent Talents
in University (NCET) by Ministry of Education of P. R. China (Grant
No. NCET-13-0991).



\end{document}